\documentclass[aps,prl,floats,twocolumn,twoside,floatfix]{revtex4}

\usepackage{graphicx}
\usepackage{amsmath,amssymb}
\usepackage{psfrag}
\usepackage[utf8]{inputenc} 

\begin{document}

\title{Dynamic facilitation picture of a higher-order glass
  singularity} 

\author{Mauro Sellitto$^1$, Daniele De Martino$^2$, Fabio
  Caccioli$^2$, Jeferson J.  Arenzon$^3$}

\affiliation{$^1$United World College in Mostar,
  \v{S}panski trg 1, 88000 Mostar, Bosnia and Herzegovina \\
  $^2$International School for Advanced Studies, via Bonomea 265,
  I-31136 Trieste, Italy \\
  $^3$Instituto de F{\'\i}sica, Universidade Federal do Rio Grande do
  Sul, CP 15051, 91501-970 Porto Alegre RS, Brazil}

\newcommand{\tw}{t_{\scriptstyle \rm w}}
\newcommand{\kB}{k_{\scriptscriptstyle \rm B}}
\newcommand{\Tc}{T_{\scriptstyle \rm c}}
\newcommand{\Tg}{T_{\scriptstyle \rm c}}
\newcommand{\pc}{p_{\scriptstyle \rm c}}
\newcommand{\qc}{q_{\scriptstyle \rm c}}
\newcommand{\Phic}{\Phi_{\scriptstyle \rm c}}

\begin{abstract}
  We show that facilitated spin mixtures with a tunable facilitation
  reproduce, on a Bethe lattice, the simplest higher-order singularity
  scenario predicted by the mode-coupling theory (MCT) of liquid-glass
  transition.  Depending on the facilitation strength, they yield
  either a discontinuous glass transition or a continuous one, with no
  underlying thermodynamic singularity.  Similar results are obtained
  for facilitated spin models on a diluted Bethe lattice.  The
  mechanism of dynamical arrest in these systems can be interpreted in
  terms of bootstrap and standard percolation and corresponds to a
  crossover from a compact to a fractal structure of the incipient
  spanning cluster of frozen spins. Theoretical and numerical
  simulation results are fully consistent with MCT predictions.
\end{abstract}


\maketitle

Although the glassy state of matter has been a long time fascinating
topic for physicists, it still presents several mysterious
aspects~\cite{Langer07}.  Perhaps, the most controversial one is the
very nature of the glass transition: that is the question of whether
vitrification is a purely dynamical process or rather the (dynamical)
manifestation of a genuine thermodynamic amorphous phase.
In spite of many progresses (for reviews,
see~\cite{Gotze09,BiKo05,BiBo09,RiSo03,GaCh10}) the
problem remains widely open. On one hand, there are notorious
experimental and computational difficulties related to the exceedingly
long equilibration times of macroscopic samples, and to the
possibility of detecting unambiguously the elusive `amorphous order'.
On the other hand, theoretical modelling of glassy systems has made
clear that slow relaxation processes are ubiquitous and may result
from very distinct mechanisms.

In a situation in which it is unknown how much intertwined
thermodynamics and dynamics are, it would be particularly advantageous
from a methodological point of view to identify those peculiar glassy
features that can be reproduced with no reference to specific
energetic interactions.  In this respect, facilitated spin models
first introduced by Fredrickson and Andersen are particularly
useful~\cite{FrAn84}, as they are constructed in such a way to have a
manifestly uninteresting thermodynamics. Hence they allow to clearly
disentangle, albeit in an arguably artificial way, dynamic aspects
from static ones.

The first assumption of the dynamic facilitation approach is that, on
a suitable coarse-grained lengthscale, one can model the structure of
a liquid by an assembly of high/low density mesoscopic cells which
have {\em no} static interaction. Binary spin variables, taking on
value $\pm 1$, can be simply assigned to these cells. The next crucial
step is to postulate that there exists a timescale over which the
effective microscopic dynamics takes a deceptively simple form: local
changes in cells structure occur if and only if there is a
sufficiently large number, say $f$, of nearby low-density cells ($f$
is called facilitation parameter).  The latter assumption is actually
difficult to derive by analytical means. Nevertheless, it can be
justified on physical grounds: it mimics the cage effect and gives
arise to a variety of remarkable, and sometimes unexpected,
glassy features, even if the thermodynamics is completely
trivial~\cite{RiSo03}.
Although this line of research has been pursued very actively in
recent years, little attention has been paid to the possibility of
reproducing, within this framework, more complex types of glassy
behaviour (for some exceptions, see~\cite{GeRe05,MoCo06}). In fact,
even simple schematic MCT models predict the occurrence of
topologically stable singularities of higher
complexity~\cite{Gotze09}.  Liquids confined in a disordered porous
matrix~\cite{Krako} and attractive colloids~\cite{Dawson,Pham} are
some examples in which these scenarios have been recently observed.
Solvable microscopic realizations of such systems would be highly
valuable both for a deeper understanding of complex glassy features
and for clarifying the limits of MCT.  This is quite a delicate issue,
however, as mean-field disordered systems which are supposed to be
exactly described by MCT are spoiled by finite-size effects and
pre-asymptotic corrections so strong to prevent a direct observation
of MCT predictions~\cite{BrKoBi,SaBiBiBo}. For this reason, we follow
here an alternative route.

Motivated by recent findings on the annealed-quenched
mixtures~\cite{Krako}, we have generalized the dynamic facilitation
approach by allowing for an inhomogeneous distribution of facilitation
or, equivalently, as we shall see, of lattice connectivity. While the
latter may originate from a geometrically disordered environment,
e.g. a porous matrix, the former can be thought of as resulting from
the coexistence of different lengthscales in the problem,
e.g. mixtures of more or less mobile molecules/polymers with small and
large size.
By doing so, we show that facilitated spin models on Bethe lattice
provide a close microscopic realization of the simplest higher-order
bifurcation singularity scenario predicted by MCT, in which two
liquid-glass transition lines (of types A and B) join smoothly at a
common endpoint.

%\paragraph{The model.}

{\it The model.} Facilitated spin models consist of $N$ non-interacting spins
$\sigma_i=\pm 1$, $i=1,\dots,N$ with Hamiltonian ${\cal H} = -
\frac{h}{2} \sum_{i=1}^N \sigma_i$, evolving with a Metropolis-like
dynamics: at each time step a randomly chosen spin is flipped with
transition probability: $w(\sigma_i \to -\sigma_i) = {\rm min} \left\{
  1, {\rm e}^{- h \sigma_i/\kB T} \right\}$, if and only if at least
$f$ of its $z$ neighboring spins are in the state $-1$ (hereafter
$h/\kB=1$).  
On a Bethe lattice, the dynamics can be characterized by exploiting
the relation with the bootstrap percolation~\cite{ChLeRe79,SeBiTo05}.
Disregarding the less interesting noncooperative case, $f=1$, in which
there is no transition, one can distinguish two cases.  {\em i)} For
$z-1>f>1$, the system undergoes a dynamical arrest: below a certain
temperature $\Tg$, the fraction of frozen spins $\Phi$, which plays
the role of the non-ergodicity parameter in MCT, jumps from zero to a
finite value. This corresponds to the sudden emergence of a giant
cluster of frozen spins (i.e. spins that are surrounded by more than
$f$ neighbouring 1 spins) with compact structure.  This dynamic
transition has a very peculiar {\em hybrid} nature: it is
discontinuous and, at the same time, has diverging fluctuations as in
continuous phase transitions.  The geometric origin of this behaviour
has been understood quite in detail as being related to the divergence
of the size of {\em corona} clusters near the
transition~\cite{DoGoMe06,ScLiCh06}.
Several results~\cite{SeBiTo05}, including those related to large
scale cooperative rearrangements responsible for slow
dynamics~\cite{MoSe05}, have suggested a strong analogy
with MCT.
{\em ii)} For $f=z-1,z$ the transition is continuous as bootstrap
percolation is equivalent to conventional percolation~\cite{ChLeRe79},
and for this reason it has attracted less interest. The facilitated
spin dynamics of these systems has never been explored to our
knowledge.
To study the crossover between integer values of $f$ we introduce a
new facilitated spin model in which the facilitation strength can be
continuously tuned. This is obtained by making the facilitation
parameter a lattice site dependent quenched random variable, in close
analogy with the bootstrap percolation problem studied by
Branco~\cite{Branco93}.  For concreteness, we shall consider a Bethe
lattice with coordination number $z=4$, in which the facilitation
$f_i$ is chosen from the probability distribution:
\begin{equation}
  P(f_i) =  (1-q)\ \delta_{f_i,2} + (q-r) \ \delta_{f_i,3}+ r \ \delta_{f_i,4} , 
\label{eq.distf}
\end{equation}
with $ 1 \ge q \ge r \ge 0$. By tuning $q$ and $r$ we can thus explore
the discontinuous/continuous glass transition crossover.

\begin{figure}[htbp]
\includegraphics[width=8.5cm]{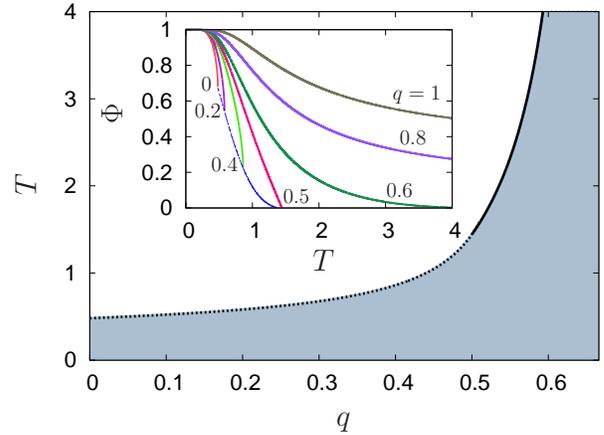}
\caption{Phase diagram for $z=4$ and facilitation as in
  (\ref{eq.distf}). The dark region is the glassy phase. The
  dashed/solid line is the discontinuous/continuous transition.  Inset:
  Fraction of frozen spins vs temperature for several values of $q$
  and $r=10^{-3}$. Below $q=1/2$, $\Phi$ jumps to a finite value at
  the transition which is represented by the dotted line.}
\label{fig.diag}
\end{figure}

{\it Exact results.} The tree-like structure of the Bethe lattice
allows for an exact calculation of the phase diagram.  Following
Refs.~\cite{ChLeRe79,Branco93,SeBiTo05} we get:
\begin{equation}
  \Tg(q) = \left\{
\begin{array}{l}
  \displaystyle \frac{1}{\log\left( 8-12q \right)}, 
  \text{   if  } 0 \leq q \leq 1/2 \\
  \displaystyle - \frac{1}{\log\left( 3q-1 \right)}, 
  \text{   if  } 1/2 < q < 2/3 . 
\end{array}
\right. 
\label{eq.pc}
\end{equation}
The phase diagram is depicted in fig.~\ref{fig.diag}, and comprises
two lines that smoothly join at $q=1/2$. As expected, the introduction
of less facilitated spins increases the glass transition temperature,
as compared to the pure case, $q=0$~\cite{SeBiTo05}.  The precise
nature of the two glass transitions depends on the behaviour of $\Phi$
near $\Tg$.  Denoting with $p=1/(1+{\rm e}^{-1/T})$ the probability
that a spin is 1 in thermal equilibrium, we find
\begin{eqnarray}
\Phi &=& p\left[ x^3(4-3x)+6qx^2(1-x)^2+4r x (1-x)^3 \right] \label{eq.phi}\\
     &+& (1-p)\left[ y^3(4-3y)+6qy^2(1-y)^2+4 r y(1-y)^3 \right]\nonumber
\end{eqnarray}
where $ y = p\left[ x^3 + 3(1-q)x^2(1-x) + 3r x (1-x)^2\right] $, and $x
= \left[3p(1-2q)+\sqrt{9p^2-8p+12qp(1-p) } \right]/2p(2-3q)$.
The behaviour of $\Phi$ as a function of the temperature is shown in
the inset of Fig.~\ref{fig.diag}, for several values of $q$ and with
$r=10^{-3}$.
\begin{figure}[hbtp]
  \includegraphics[width=8.cm]{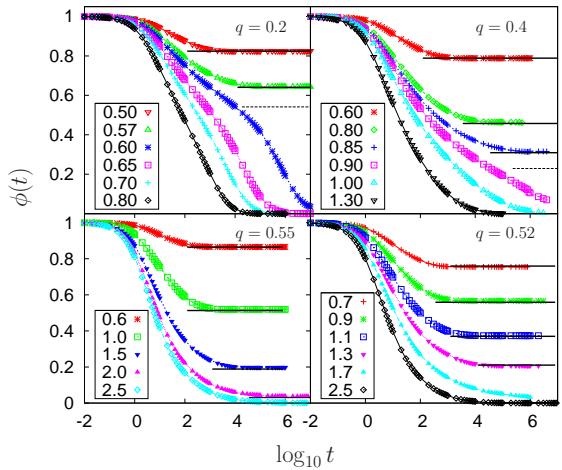}
  \caption{Persistence vs time for several temperatures, indicated in
    the key, and values of $q$. The transition is
    discontinuous/continuous for $q$ smaller/larger than 1/2. We use
    $r=10^{-3}$, except for $q=0.52$ where $r=0$. The horizontal lines
    show the theoretical prediction for the plateau heights (the
    dashed one is the critical plateau at $T=\Tc$). The simulations
    are performed on lattices with $N=10^5$ to $5\times 10^5$
    sites and averaged over 2-20 samples.}
  \label{fig.pers_f2f3f4}
\end{figure}
Notice that $\Phi$ generally depends on $r$ while the phase diagram
does not. For $ 0 \leq q \leq 1/2$, $\Phi$ jumps to a finite value
$\Phic=\Phi(\Tg)$ on the transition line, meaning that the infinite
cluster of frozen spins has a compact structure. This structure turns
out to be quite resilient against random inhomogeneities in the
facilitation strength, in a rather large range of $q$.
The critical exponent $\beta$ associated to the order parameter $\Phi$
is obtained by expanding the above equations in the small parameter
$\epsilon=T-\Tg$.
As expected, we find $\Phi-\Phic \sim \epsilon^{\beta}$, with
$\beta=1/2$, that is the typical square-root dependence well known in
MCT and in other systems with hybrid transition.
For $1/2 < q < 2/3$ the glass transition changes nature as $\Phi$
departs smoothly from zero with a power-law behaviour $\Phi \sim
\epsilon^{\beta}$, and the correspondence with standard percolation
suggests that the giant cluster of frozen spins has a fractal
structure. Interestingly, in this case $\beta$ depends on $r$: for
$r=0$ one has $\beta=2$, while as soon as a negligibly small amount of
spins with $f_i=4$ is introduced into the system, $r>0$, one has
$\beta=1$.  The robustness of the latter behaviour reflects the fact
that the mass of the fractal cluster of frozen spins is essentially
dominated by the {\em dangling ends}, that is those parts of the
cluster which are connected to the backbone by a single site.  Only
when $r=0$, the latters are completely removed from the infinite
cluster and the exponent changes to $\beta=2$~\cite{Branco93}.
Hence, the general scenario emerging for an arbitrary ternary mixture
is that of two distinct glass transitions with $\beta=1/2$ (for the
discontinuous case), and $\beta=1$ (for the continuous one).  These critical
exponents reproduce exactly the MCT results for the $F_{12}$ schematic
model~\cite{Gotze09}, in which the memory kernel takes the form
$m_{12} = v_1 \phi(t) + v_2 \phi^2(t)$.

{\it Numerical simulations.} We now turn to numerical simulation to
explore those features of equilibrium relaxation which are relevant
for a comparison with MCT.
Testing MCT in systems which are described by a somewhat ad hoc kinetic
rule is particularly interesting because it allows us to probe the
degree of universality of MCT results beyond the context (the actual
Newtonian or Brownian liquid dynamics) and the approximations in which
they were originally derived.
The dynamics of facilitated spin systems is conveniently characterized
by the persistence $\phi(t)$, i.e. the probability that a spin has
never flipped between times $0$ and $t$. The long-time limit of
$\phi(t)$, which plays the role of the Edwards-Anderson parameter in
spin-glasses, is directly related to the fraction of frozen spins
$\Phi = \lim_{t \to \infty} \phi(t)$, which is known to describe
broken ergodicity.
\begin{figure}[htbp]
\includegraphics[width=8.5cm]{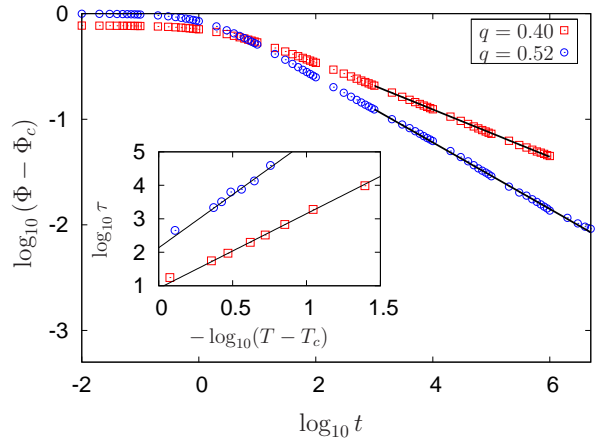}
\caption{Equilibrium relaxation at criticality for discontinuous ($q=0.4$)
  and continuous ($q=0.52$) glass transition. Solid lines are
  power-law fits with exponent $a=0.224$ and $a=0.315$. Inset:
  $\beta$-relaxation time $\tau$ vs temperature $T$. Solid lines are
  power-laws with exponent $1/2a$ and $1/a$, respectively.}
\label{fig.pers_f2f3f4_Tc}
\end{figure}
Simulation results for the persistence are shown in
fig.~\ref{fig.pers_f2f3f4} for various $q$, above and
below the glass transition temperature $\Tg(q)$. To avoid dynamic
reducibility problems the ternary mixture has a negligibly small
fraction of spins with $f=4$, typically $r=10^{-3}$.
We see that inside the glassy phase, the persistence attains a finite
plateau which is in excellent agreement with the value of $\Phi$
analytically computed in the previous section.
On approaching $\Tg(q)$ from above and for $q<1/2$ we find typical
signatures of MCT: two-step decay of $\phi(t)$, late stage stretched
exponential relaxation, and power-law form of equilibration time. When
$q \to 1/2$ (and so $\Phic \to 0$), the range over which the
time-temperature superposition principle holds shrinks, and
consequently no simple scaling form of relaxation data at various
temperature can be found in this limit.

To make a more stringent test of MCT we now examine in detail the
critical dynamics on the glass transition lines, for which MCT
predicts distinctive patterns of universal critical behaviour.
When the system is relaxing exactly at the critical temperature
$\Tg(q)$, MCT predicts the power law decay, $\phi(t)-\phi(\infty) \sim
t^{-a(q)}$, irrespective of whether the long-time limit of persistence
is finite or zero.
Fig.~\ref{fig.pers_f2f3f4_Tc} shows that indeed this power
law behaviour is very well obeyed.
A further remarkable prediction of MCT is that the exponent $a(q)$ is
intimately related, in a way that depends on the nature of the glass
transition, to the power-law exponent describing the divergence of the
characteristic time $\tau$ on approaching the plateau at $\Phic$ (the
so-called $\beta$-relaxation regime): when temperature gets closer to
$\Tg(q)$ one has $\tau \sim \epsilon^{-1/2a(q)}$ for the discontinuous
transition, and $\tau \sim \epsilon^{-1/a(q)}$ for the continuous one.
These two distinct behaviors are tested in the inset of
Fig.~\ref{fig.pers_f2f3f4_Tc} by assuming as a reasonable definition
of the $\beta$-relaxation time, $\tau$, the integral of $\phi(t)$ from
zero up to the time $t^*$ such that $\phi(t^*) = \Phic$.
The estimated $\tau$ behaves as a power-law of $\epsilon$
near $\Tg$ and, more importantly, the exponents perfectly agree with
MCT, for both the discontinuous and continuous glass transition.
The departure from the plateau, the so-called $\alpha$-relaxation
regime, is also consistent with the exponent $b$ obtained from the
universal MCT relation $ \Gamma(1-a)^2/\Gamma(1-2a) =
\Gamma(1+b)^2/\Gamma(1+2b)$, though this regime is much more difficult
to analyse due the exceedingly long time scales of numerical
simulations.

The excellent agreement we found with MCT could not be obviously
anticipated by looking at the `simple' facilitation rule and is all the
more remarkable by considering that {\it i)} no exact mode-coupling
equation holds for facilitated spin models~\cite{PiYoAn00}, and {\it
  ii)} the facilitation rule is admittedly quite remote from the
actual microscopic liquid dynamics.  Furthermore, our results imply
that no thermodynamic transition is actually needed to observe a
complex glassy scenario, and suggest that the difficulty to observe
MCT predictions in mean-field spin-glasses, is generally related to
the fact that the dynamic and static transition properties in these
systems are tightly twisted.

{\it Conclusions.}  To summarize, we have extended the dynamic
facilitation approach to glassy systems by introducing facilitated
spin mixtures which exhibit a crossover from a discontinuous to a continuous
glass transition, and shown that they provide a close microscopic
realization of the simplest higher-order glass singularity predicted
by MCT.
Although we focused on Bethe lattice with fixed connectivity and
random distribution of facilitation, it is worth to remark that one
can consider an equivalent variant, in which the Bethe lattice is
diluted and the facilitation is uniform.  For example, results
qualitatively similar to those reported above can be obtained when the
local lattice connectivity $z_i$ is distributed according to
$P(z_i)=(1-q)\delta_{z_i,4} + (q-r)\delta_{z_i,3} + r\delta_{z_i,2}$,
and $f_i=2$ on every site. In this case, less connected sites turn out
to have a smaller probability to flip, just as if they were less
facilitated. We thus expect, and indeed find, that both variants give
essentially the same results.
Thus, the crossover between the two glass transitions is obtained by
varying either the local connectivity or the facilitation strength,
and corresponds to a passage from bootstrap to standard percolation
transition.
Consistently with MCT predictions, the critical dynamics on both glass
transition lines is characterized by power-law decays with exponents
closely related to those describing the divergence of the
$\beta$-relaxation time.
We also mention that when a fraction of spins is overfacilitated as
compared to the case $f_i=2$, i.e. when $ P(f_i) = (1-q)\delta_{f_i,2}
+ (q-r)\delta_{f_i,1}+ r \delta_{f_i,0} $, there is only a
discontinuous glass transition line.

Our work can be naturally developed in several directions.  The most
obvious one consists in considering kinetically constrained particle
mixtures, what would be a much closer microscopic realization of a
fluid confined in a disordered porous matrix.
The comparison with geometrically constrained lattice glass
mixtures~\cite{DaReBi}, would then allow for a better understanding of
the distinguishing features of glassy systems with and without an
ideal Kauzmann transition.
We then expect that introducing a suitable static attraction~\cite{GeRe05}
can account for nonmonotonic dependence of relaxation time on
attraction strength and, possibly, for phase reentrant behaviour.
Finally, the existence of kinetic models with a glass transition in
$d=2$~\cite{ToBiFi06}, suggests that there is no apparent limitation
to extending our approach to finite dimensions.

\acknowledgments MS warmly thanks Prof. G\"otze for kindly clarifying
some issues related to MCT, and the brazilian agency CAPES for a PVE
grant.  DDM and FC acknowledge MIUR for the grant 2007JHLPEZ.  JJA is
a member of the INCT-Sistemas Complexos and is partially supported by
the brazilian agencies CNPq (Prosul-490440/2007) and FAPERGS.

\end{document}